\documentclass[aps,prl,twocolumn,showpacs]{revtex4}
\usepackage{graphics}
\usepackage{graphicx}

\begin{document}

\title{Ionization fronts in negative corona discharges.}
\author{Manuel Array\'as, Marco A. Fontelos and Jos\'e L. Trueba}
\address{Departamento de Matem\'aticas y F\'{\i}sica Aplicada, Universidad Rey Juan Carlos, Tulip\'{a}n s/n, 28933 M\'{o}stoles, Madrid, Spain}

\begin{abstract}
In this paper we use a hydrodynamic minimal streamer model
to study negative corona discharge. By reformulating the
model in terms of a quantity called shielding factor, we
deduce laws for the evolution in time of both the radius
and the intensity of ionization fronts. We also      
compute the evolution of the front thickness under the
conditions for which it diffuses due to the geometry of the
problem and show its self-similar character.
\end{abstract}

\date{\today}
 \maketitle


A common feature in transient discharges preceding dielectric breakdown
is the creation of a non-equilibrium plasma through the propagation of a
nonlinear ionization wave into a previously non-ionized region.
Modern concepts of pattern formation which have been already applied in different contexts (see e.g \cite{Buck} and \cite{Meron}), have also been used in order to gain new analytical insight in this old problem \cite{Ute}.

When a sufficiently high voltage is suddenly applied
to a medium with low or vanishing conductivity, extending fingers of
ionized matter develop. They are called streamers and
are ubiquitous in nature and technology \cite{Rai,Eddie}. A minimal
streamer model, consisting of a fluid approximation with local field-dependent impact ionization reaction
in a non-attaching gas like argon or nitrogen, has been used to study
the basics of streamer dynamics \cite{DW,Vit,Ute,ME,ME1,Andrea}.
The essential properties of planar fronts for this minimal model have been obtained as a first step towards an effective interface description. For the
planar fronts, the mechanism of dynamical front selection has been
understood and explained \cite{pulled1,pulled2}. The dispersion relation for transversal Fourier-modes of planar negative ionization fronts has been derived \cite{ME1} in an attempt to study how the electric screening layer influences their stability.

As a further step, we consider in this paper the evolution of negative ionization fronts in a gas under the influence of a non-uniform external electric field. The field is created by a potential difference $V_{0}$ applied between an electrodes pair. The geometry of the electrodes determines the non-uniformity of the electric field. Discharge develops in the high field region near the sharper electrode and it spreads out towards the other electrode. This type of discharges is called corona. It is a negative corona discharge when the electrode with the strongest curvature is connected to the negative terminal of the power supply. We will consider this case.

 The so-called minimum model consists in the following dimensionless set of equations (for physical parameters and dimensional analysis, we refer to
the previous discussions in \cite{Ute,ME,ME1,Andrea}):
\begin{eqnarray}
\frac{\partial n_e}{\partial t} - \nabla \cdot {\bf j} = n_e f({\bf E}), \label{elect} \\
\frac{\partial n_i}{\partial t} = n_e f({\bf E}), \label{ion} \\
\nabla \cdot {\bf E} = n_i - n_e . \label{gauss}
\end{eqnarray}
The equation (\ref{elect}) describes the rate of change of the local dimensionless electron density $n_e$. It is equal to the
divergence of the local electron current density ${\bf j}$ plus a source term $n_e f({\bf E})$ representing the
generation of electron-ion pairs due to the impact of accelerated electrons onto neutral molecules of gas. The
value of $f({\bf E})$ is given by the Townsend approximation
\begin{equation}
  \label{townsend}
  f({\bf E}) = E \exp{(-1/E)},
\end{equation}
\noindent
where $E$ is the modulus of the local electric field ${\bf E}$. In equation (\ref{ion}) we consider that the rate of
change of the ion density $n_i$ is equal to the source term due to impact, since we take the ion current density to
be negligible in a first approximation (the speed of ions is typically much smaller than that of electrons). The
local value of the electron current density is specified as
\begin{equation}
  \label{current}
  {\bf j} = n_e {\bf E} + D \nabla n_e, \label{current}
\end{equation}
\noindent
 using Ohm's law in the
first term and considering diffusion effects in the second one. Note that this expression does not include the
effect of the magnetic field created by the motion of electrons, as it is supposed that their speed is much smaller
than the speed of light. Equation (\ref{gauss}) is Gauss' law in local form, coupling the electric field to the charge densities.

Since our primary goal in this paper is to address the effects of curvature in front propagation, we will neglect diffusion effects as in \cite{ME1}. That allows us to reduce the set of equations (\ref{elect})-(\ref{gauss}) into a simpler form in order to give analytical
results for the evolution of the ionization fronts. From (\ref{elect}),
(\ref{ion}), and (\ref{current}) with $D=0$, we obtain
\begin{equation}
\frac{\partial}{\partial t} \left( n_i - n_e \right) + \nabla
\cdot \left( n_e {\bf E} \right) = 0, \label{paso1}
\end{equation}
and from (\ref{gauss}), taking the time derivative,
\begin{equation}
\nabla \cdot \left( \frac{\partial {\bf E}}{\partial t} \right) =
\frac{\partial}{\partial t} \left( n_i - n_e \right).
\label{paso2}
\end{equation}
Equations (\ref{paso1}) and (\ref{paso2}) give then
\begin{equation}
\nabla \cdot \left( \frac{\partial {\bf E}}{\partial t} + n_e
{\bf E} \right) =0. \label{paso3}
\end{equation}
The term inside brackets in (\ref{paso3}) is, due to Maxwell equations, proportional to the curl of the magnetic
field in the gas. As it is supposed that the magnetic field is negligible, we can take it equal to zero and integrate in time yielding
\begin{equation}
{\bf E} ({\bf r}, t) = {\bf E}_{0} ({\bf r}) \exp{ \left( -
\int_{0}^{t} d\tau n_e ({\bf r}, \tau ) \right) },
\label{paso5}
\end{equation}
which gives the local electric field ${\bf E}$ in terms of the
initial electric field ${\bf E}_{0}$ and the electron density
$n_e$ integrated in time. Equation (\ref{paso5}) motivates the definition of the quantity
\begin{equation}
u({\bf r},t) =\exp{ \left( - \int_{0}^{t} d\tau n_e ({\bf
r},
\tau ) \right) }. \label{defu}
\end{equation}
If this quantity is completely determined in a particular problem,
then using equations (\ref{paso5}), (\ref{defu}) and (\ref{gauss}) all the physical fields can be obtained through the
expressions
\begin{eqnarray}
{\bf E} ({\bf r}, t) &=& {\bf E}_{0} ({\bf r}) u({\bf r},t),
\label{relEu} \\
n_e ({\bf r}, t) &=& -\frac{1}{u({\bf r},t)} \frac{\partial
u({\bf r},t)}{\partial t}, \label{relsigmau} \\
n_i ({\bf r}, t) &=& -\frac{1}{u({\bf r},t)} \frac{\partial
u({\bf r},t)}{\partial t} \nonumber\\ &+& \nabla \cdot \left( {\bf E}_{0} ({\bf
r}) u({\bf r},t) \right), \label{relrhou}
\end{eqnarray}
in which the initial condition ${\bf E}_{0}$ for the electric field should be known.  Equation (\ref{relEu})
reveals clearly the role played by the function $u({\bf r},t)$ as a factor modulating the electric field ${\bf E}$ at any
time. The electronic density is positive so $u({\bf r},t)$ decays damping the electric field. For this reason we call it the {\it shielding factor}. The shielding factor determines a screening length which changes with time: a kind of Debye's length which moves with the front leaving behind a neutral plasma.

The problem is thus reduced to finding equations and conditions for
the shielding factor $u({\bf r},t)$ from equations and conditions for the
physical quantities ${\bf E}$, $n_e$ and $n_i$. Substituting (\ref{relEu})-(\ref{relrhou}) into the model equations (\ref{elect})-(\ref{gauss}), after some algebraic manipulations and integrating once in time, the evolution of $u({\bf r},t)$ is given by
\begin{eqnarray}
\frac{1}{u} \frac{\partial u}{\partial t} = \nabla \cdot \left(
{\bf E}_{0} u \right)  - n_{i0} ({\bf r}) -
\int_{E_{0}u}^{E_{0}} \exp{\left( \frac{-1}{s} \right)} d s ,
\label{evolufinal1} \\
u ({\bf r},0) = u_{0} ({\bf r}) = 1,\;\;\;\;\;\;\;\;\; \label{evolufinal2}
\end{eqnarray}
where $E_{0}$ is the modulus of ${\bf E}_{0}$ and $n_{i0}$ is the initial ion density. Boundary conditions should be imposed depending on the particular physical situation.

In what follows we will consider a typical corona geometry: two spherical plates with internal radius $R_{0}$ and
$R_{1}>>R_{0}$, respectively. An electric potential difference $V_{0}$ is applied to these plates, so that
$V(R_{1})-V(R_{0})=V_{0}> 0$. The initial seed of ionization is
taken to be neutral so that
\begin{equation}
n_{e0}(r) = n_{i0} (r) = \rho_{0}(r).
\label{icneutral|}
\end{equation}
We consider the evolution of negative ionization fronts towards the positive plate
at $r=R_{1}$. The initial electric field ${\bf E}_{0}(r)$ between the plates is
\begin{equation}
{\bf E}_{0} (r) = -\frac{C}{r^2} {\bf u}_{r}, \; C = V_{0}
\frac{R_{0} R_{1}}{R_{1}-R_{0}}. \label{spher2}
\end{equation}
We substitute (\ref{spher2}) into equation (\ref{evolufinal1}) and
change the spatial variable $r$ to
\begin{equation}
x=\frac{r^3}{3C}, \label{spher8}
\end{equation}
so that the evolution for the screening factor takes the form
\begin{equation}
\frac{\partial u}{\partial t} + u \frac{\partial u}{\partial x} = - u \rho_{0} (x) - u \int_{u\left(\frac{C}{9 x^2}\right)^{1/3}
}^{\left(\frac{C}{9 x^2}\right)^{1/3}} \exp{\left( \frac{-1}{s} \right)} d s . \label{spher9}
\end{equation}
The equation (\ref{spher9}) governing the behaviour of the screening factor is a Burgers' type equation, where $\rho_{0}(x)$ is the initial distribution of charge. The condition for the initial value of the screening factor is, by (\ref{evolufinal2}), $u(x,0)=1$. Following the usual procedure of resolution of Burgers' equation we can integrate equation (\ref{spher9}) along the characteristics $x_c(t)$ defined by
\begin{equation}
\frac{dx_c(t)}{dt}=u(x_c(t),t),
\label{eq:char}
\end{equation}
transforming (\ref{spher9}) into an ordinary differential equation.

First the case of sufficiently localized initial conditions is considered. More specifically, the initial electron density strictly vanishes beyond a certain point. Under similar conditions, the existence of shock fronts with constant velocity has been predicted for the simpler planar geometry \cite{Ute}.

\begin{figure}
\begin{center}
\includegraphics[width=0.49\textwidth,height=0.35\textwidth]{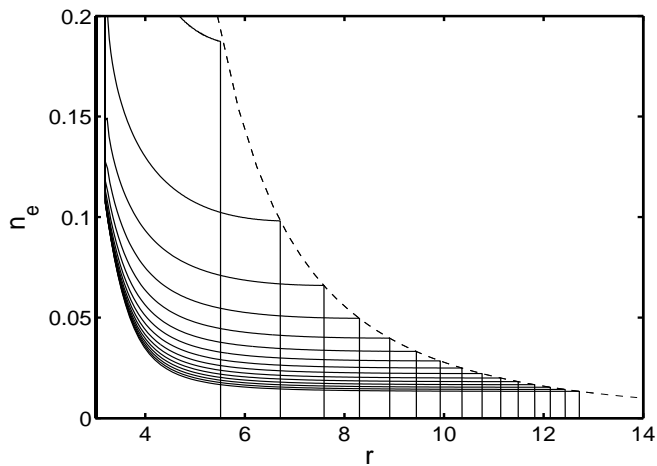}
\end{center}
\caption{The shock wave development at regular intervals of time when the initial charge distribution is well localized. The amplitude of the front calculated analytically is plotted in dashed line.}
\label{fig1}
\end{figure}

Taking a homogeneous thin layer of width $\delta <<
(R_{1}-R_{0})$ from $r=R_{0}$ to $r=R_{0}+\delta$, the initial
charge distribution is then
\begin{eqnarray}
n_{e0}(r) = n_{i0} (r) &=& \rho_{0},\; \; R_{0}<r< R_{0}+\delta, \nonumber \\
n_{e0}(r) = n_{i0} (r) &=& 0 , \;\; R_{0}+\delta < r < R_{1}.
\label{spher1}
\end{eqnarray}
In figure \ref{fig1} we show the electron density distribution $n_e$ which corresponds to some arbitrary choice of parameters $\rho_{0}, \delta, R_{0}, R_{1}, V_0$. The electron density has been calculated using expression (\ref{relsigmau}) and plotted as a function of $r$ at different times $t$. There appears a sharp shock with decaying amplitude, separating the region with charge and the region without charge.

\begin{figure}
\begin{center}
\includegraphics[width=0.49\textwidth,height=0.35\textwidth]{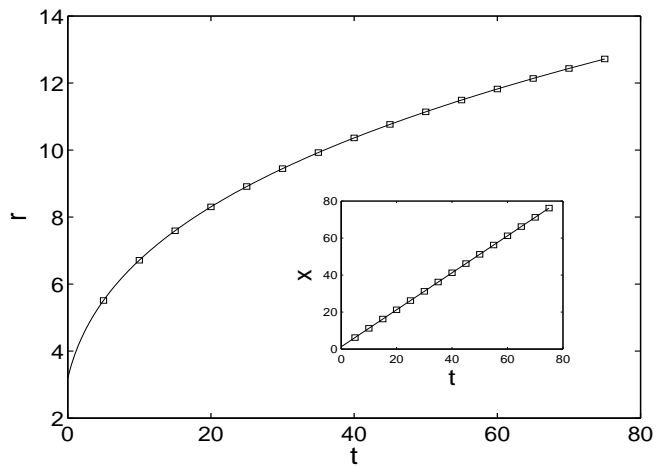}
\end{center}
\caption{Position of the shock front versus time. Squares are numerical results and the continuous lines are the theoretical predictions. In the inset it is shown the position in the scaled variable $x$. It can be observed the linear dependence in this variable as predicted in (\ref{a333}).}
\label{fig2}
\end{figure}

 From these numerical data, we can measure the velocity of propagation of such front. In figure \ref{fig2}, it is plotted the position of the shock $r_f$ as a function of time. The velocity of propagation is clearly not constant. However, if we plot the position of the front in terms of $x$, one can observe the following linear relation (see inset figure \ref{fig2})
\begin{equation}
x_f(t)=t+x_{0},
\label{eq:linear}
\end{equation}
which implies, in terms of the original variable $r$, an asymptotic behaviour
\begin{equation}
r_f(t)\sim(3C)^{1/3}\;t^{1/3}
\label{eq:nolinear}
\end{equation}
 for the position of the front.

Remarkably we can deduce expressions for both the amplitude and propagation velocity of the shock in explicit analytical form. In order to do that, we write locally near the front the solution as
\begin{equation}
u(x,t)=1-a(t)\varphi(\xi).\;\; \xi= x-x_f(t) \label{a25}
\end{equation}
We substitute this ansatz into (\ref{spher9}) and since the integral term is small when $x >> 1$ we get
\begin{equation}
a(t)\varphi^{\prime}(\xi)-a(t)\varphi^{\prime}(\xi)x^{\prime}_f(t)-a^{\prime }(t)\varphi(\xi)+a^{2}(t)\varphi (\xi)\varphi^{\prime
}(\xi)\approx 0, \label{a27}
\end{equation}
implying that
\begin{eqnarray}
x_f(t)&=&t+x_0\;,\label{a333} \\
a(t)&=& \frac{\beta}{(t+t_{0})}, \label{a30} \\
\varphi (\xi) &=& {\beta}^{-1}(x-t-x_{0}), \label{a31}
\end{eqnarray}
where $\beta$ is an arbitrary constant to be fixed from initial conditions.
Equation (\ref{a333}) proves that the position of the shock front follows the law (\ref{eq:nolinear}) for spherical geometry.

From (\ref{relsigmau}), the electron density reads
\begin{equation}
n_e(x,t)\approx\left\{\begin{array}{ll}\frac{1}{t+t_0}\;\frac{1+
(x-t-x_0)/(t+t_0)}{1-
(x-t-x_0)/(t+t_0)} &,\; x\le t+x_0\\
0 \; &,\; x>t+x_0
\end{array}
\right. ,
\end{equation}
which implies that the amplitude of the front decays as
\begin{equation}
  \label{eq:nee}
  n_e(x_f(t),t)=\frac{1}{t+t_0}.
\end{equation}
In figure \ref{fig1} the analytical curve (\ref{eq:nee}) has been plotted in dashed line, showing an excellent agreement with the numerical data.

We want to conclude this paper with a brief discussion of the case in which the initial charge distribution is not localized as, for instance, one such that
\begin{equation}
n_{e0}(x)=n_{i0}(x)=\rho_{0}(x)\sim e^{-\lambda x} \; ,\; x >> 1. \label{a34}
\end{equation}
For the planar case it was predicted \cite{Ute} a constant velocity for the propagation of the front, although no shock front would develop unless the decay is sufficiently fast.

\begin{figure}
\begin{center}
\includegraphics[width=0.49\textwidth,height=0.37\textwidth]{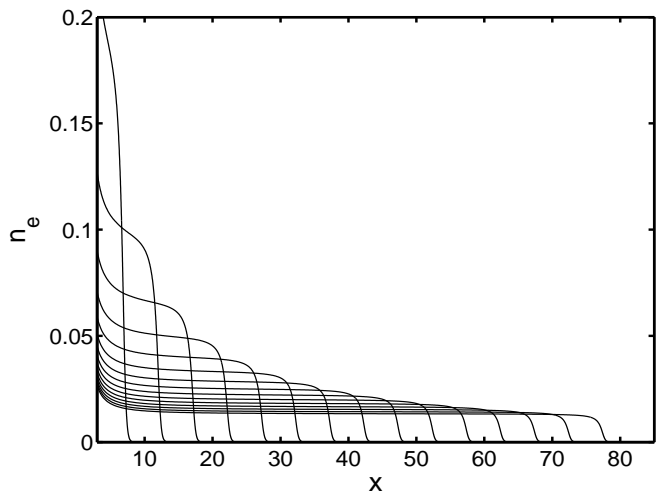}
\end{center}
\caption{Front development when the initial condition is not localized, plotted at regular intervals of time.}
 \label{fig3}
\end{figure}

As we did above, we solve the problem numerically assuming spherical symmetry, so $x$ is related to the radial coordinate $r$ by (\ref{spher8}). In figure \ref{fig3} the solution for the electron distribution is shown. The shock front does not appear in this case. Instead, a front with increasing thickness propagates. In the scaled variable $x$, we have checked that the centre of this front moves with constant velocity as the shock front does. These facts are apparent from the figure.

\begin{figure}[t]
\begin{center}
\includegraphics[width=0.49\textwidth,height=0.37\textwidth]{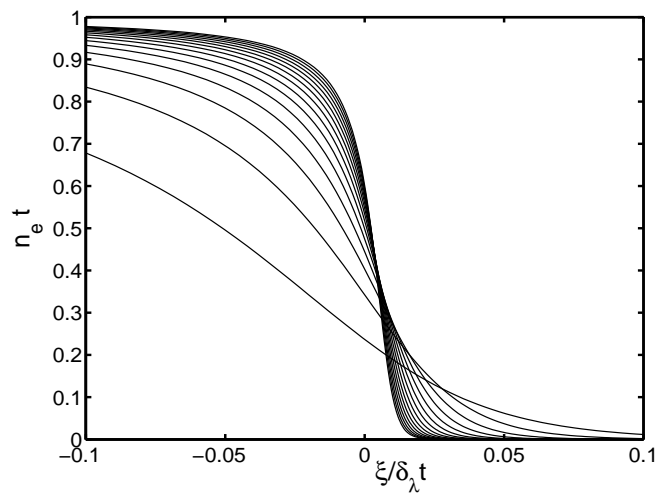}
\end{center}
\caption{Front profiles rescaled according to the self-similar law given by expression (\ref{a36}). The profiles converge asymptotically to function $f$ in that expression.}
\label{fig4}
\end{figure}

Using scaling arguments, it can be shown that the asymptotic local behaviour near the front can be described in the following self-similar form:
\begin{equation}
n_e (x,t) \approx \frac{1}{t} f\left(
\frac{\xi}{\delta_{\lambda }t}\right), \label{a36}
\end{equation}
where $\xi = x-t$, $f$ is some universal self-similar profile and $\delta_{\lambda}$ is a constant measuring the front thickness in rescaled units. Its value depends on the physical parameters and initial conditions. Hence the front presents a typical thickness
\begin{equation}
\xi_c \approx \delta _{\lambda }t . \label{a37}
\end{equation}
The fact that, even neglecting diffusion, the front spreads out linearly in time is a remarkable feature of the curved geometry considered here. For this reason it can be termed as {\it geometrical diffusion}. In figure \ref{fig4} we have plotted the numerical solutions rescaled according to (\ref{a36}) showing a clear convergence towards a universal profile.

The principal results and contributions from the work presented in this paper can be summarised as follows. First we have introduced the shielding factor as the factor damping the electric field in non-equilibrium electric discharges when the magnetic field can be considered negligible (\ref{defu}). This factor defines a characteristic length analogous to Debye's length for stationary discharges. The physics contained in the minimum model for streamer discharges can be reduced to the study of the evolution of the shielding factor.
We have derived the equation which governs its evolution (\ref{evolufinal1}) for a gas like nitrogen or argon without taking into account the diffusion of charged species and the processes of photoionization.

Then we have consider the case of a negative corona discharge with spherical symmetry. In this case, the discharge takes place in a non-homogeneous electric field and the equation for the shielding factor turns out to be a Burgers' one. We have extended the results of planar fronts to this case where the geometry is curved. Depending on initial conditions for the charge distribution, one might have negative shocks or spreading fronts. In both cases, the amplitude decreases in time and the propagation velocity follows a power law. In the case of spreading fronts we have proved the appearance of diffusion-type phenomena due to purely geometrical effects.

\end{document}